\documentclass[prd,amsmath,amssymb,superscriptaddress,twocolumn,nofootinbib,floatfix,showpacs]{revtex4}
\usepackage{epsfig}
\usepackage{graphicx}
\usepackage{bm}
\usepackage{color}
\newcommand{\newc}{\newcommand}
\def\thebiblio#1{
\begin{center}\bf \large References
\end{center}
\list
{[\arabic{enumi}]}{\settowidth\labelwidth{#1.}\leftmargin\labelwidth
 \advance\leftmargin\labelsep
 \usecounter{enumi}}
 \def\newblock{\hskip .11em plus .33em minus -.07em}
 \sloppy
 \sfcode`\.=1000\relax}

\newcommand{\eqn}[1] {Eq.~(\ref{#1})}

\def\be{\begin{equation}}
\def\ee{\end{equation}}
\def\ba{\begin{eqnarray}}
\def\ea{\end{eqnarray}}
\def \omms   {\Omega_m}
\newc{\lcdm }{$\Lambda$CDM }
\renewcommand{\(}{\left(}
\renewcommand{\)}{\right)}
\renewcommand{\[}{\left[}
\renewcommand{\]}{\right]}

\begin{document}

\title{Parametrization for the Scale Dependent Growth in Modified Gravity}

\author{Juan C. Bueno Sanchez}\email{jbueno@cc.uoi.gr}
\affiliation{Department of Physics, University of Ionnina, Greece}
\affiliation{Departamento de F\'isica At\'omica, Molecular y Nuclear, Universidad Complutense de Madrid, 28040 Madrid, Spain}

\author{James~B.~Dent} \email{jbdent@asu.edu}
\affiliation{Department of Physics and School of Earth and Space
Exploration, Arizona State University, Tempe, AZ 85287-1404}

\author{Sourish~Dutta}
\email{sourish.d@gmail.com}
\affiliation{Department of Physics and Astronomy, Vanderbilt University,
Nashville, TN 37235}

\author{Leandros Perivolaropoulos}
\email{leandros@uoi.gr}
\affiliation{Department of Physics, University of Ionnina, Greece}

\begin{abstract}
We propose a scale dependent analytic approximation to the exact linear growth of density perturbations in Scalar-Tensor (ST) cosmologies. In particular, we show that on large subhorizon scales, in the Newtonian gauge, the usual scale independent  subhorizon growth equation does not describe the growth of perturbations accurately, as a result of scale-dependent relativistic corrections to the Poisson equation. A comparison with exact linear numerical analysis indicates that our approximation is a significant improvement over the standard subhorizon scale independent result on large subhorizon scales. A comparison with the corresponding results in the Synchronous gauge demonstrates the validity and consistency of our analysis.
\end{abstract}


\maketitle

\section{Introduction}

Cosmological data from a wide range of sources including type Ia
supernovae \cite{union08, perivol, hicken}, the cosmic microwave
background \cite{Komatsu}, baryon acoustic oscillations
\cite{bao,percival}, cluster gas fractions
\cite{Samushia2007,Ettori} and gamma ray bursts
\cite{Wang,Samushia2009} seem to indicate that at least 70\% of
the energy density in the universe is in the form of an exotic,
negative-pressure component, called dark energy. While the
standard $\Lambda$CDM framework is the minimal model that
successfully accounts for observations \cite{Komatsu}, there
remain numerous viable alternatives that also pass  current
experimental tests.  These alternative models can be broadly
categorized as quintessence
\cite{RatraPeebles,Caldwell:1997ii,SteinhardtWangZlatev,WangSteinhardt:1998,dutta}
or modified gravity models
\cite{Parker:1999td,Perivolaropoulos:2005,CaldwellCoorayMelchiorri,JainZhang,NesserisPerivolaropoulos:2006,WangHuiMayHaiman,Tsujikawa,HeavensKitchingVerde,EspositoFarese:2000ij,Dvali:2000hr,Freese:2002sq,Carroll:2004de,OCallaghan:2009bu, HL, FT},
and both categories can, with sufficient tuning, replicate the
expansion history of the universe in consistence with
observations. (See \cite{Copeland, HutererTurner} for recent
reviews). In order to distinguish between these two categories of
models, it is therefore important to look beyond the expansion
rate.

The growth of structure offers a hope of breaking this degeneracy
since different growth histories can arise from models which have
similar expansion histories.  To examine the growth of structure
one examines the evolution of the linear matter density contrast
$\delta \equiv \delta\rho/\rho$ which is given in terms of the
background density $\rho$ and the perturbation $\delta\rho$. For
scales much smaller than the horizon, $\delta$ satisfies a simple
equation called the growth equation, which is scale-independent. 
Note that in what follows we use the usual definition of 'scale' as either the distance $\lambda_p$ in physical FRW coordinates or the corresponding wavenumber $k=\frac{2\pi}{\lambda_p}$.
There are many investigations which attempt to characterize the
evolution of $\delta$ through the use of a growth parameterization
which assumes a different value depending on the cosmological
model used, thereby allowing for models to be distinguished (e.g.
\cite{WangSteinhardt:1998,LinderCahn:2007,Gong:2009sp}).  The standard
definition of the growth parameter, $\gamma$, in terms of the
growth function $f$, the matter density $\Omega_m$ and the scale
factor $a$ is given as \be \label{stdpar} f(a)\equiv \frac{d
\textrm{ln} \delta}{d\textrm{ln}a} \equiv \Omega_{m}(a)^{\gamma}\,. \ee
Once this parameter is determined (for earlier theoretical
developments on the parameterization of the growth parameter and
experimental constraints on $\gamma$, see
\cite{dgp,LinderCahn:2007,Uzan:2006mf,Polarski:2007rr,Nesseris:2006er,Hawkins:2002sg,Viel:2004bf,Viel:2005ha,Kaiser:1996tp,Seo:2003pu,Mantz:2007qh,Pogosian:2005ez,Nesseris:2007pa}) one
may then be in a position to determine whether the standard
general relativistic (GR) framework of $\Lambda$CDM is responsible
for the acceleration of the universe, or some other, more exotic
process is at work.

However, in \cite{DentDutta:2008,DentDuttaPerivolaropoulos:2009},
it was demonstrated that in the Newtonian gauge (for another
interesting look at gauge issues see \cite{Wenbinlin:2001}) the
above parameterization can become inaccurate for large subhorizon
scales $\gtrsim 100h^{-1}$Mpc. In other words, if the physical
growth of structure is correctly described by the Newtonian
gauge, then it would show up as inconsistent with
scale-independent parameterization in \eqn{stdpar} and
(mistakenly) appear to be caused by exotic physics. The reason for
the discrepancy was shown to be the scale dependence of the growth
of $\delta$, which becomes important for large subhorizon scales
($\gtrsim 100h^{-1}$Mpc). An improved version of the growth
equation was derived in \cite{DentDutta:2008} which incorporates
the scale-dependence.  In \cite{DentDuttaPerivolaropoulos:2009},
a new scale-dependent parametrization of the growth function $f$
was proposed,  which was shown to account for the evolution of
$f(a)$ on these large scales with considerably greater accuracy
than \eqn{stdpar}.

In the present work, we focus on the growth of perturbations in
scalar-tensor (ST) theories of gravity. It is well known that in the sub-Hubble approximation the
growth of perturbations in these theories is also described by an
equation similar to the growth equation in GR, up to a redefinition
of the gravitational constant. Working in the Newtonian gauge,
we show that the usual growth equation approximation for $\delta$
becomes unreliable on large scales, for the same reason as in the
GR case, i.e. the effects of scale dependence. We derive an
improved version of the growth equation relevant for these models
and propose a more accurate scale-dependent parameterization for
growth.

The layout of our paper is as follows. In Section
II we discuss the growth of perturbations in
ST theories, demonstrating the failure of the usual
growth equation approximation and introducing an improved growth
equation and a new parameterization for the growth function in
these models. In Section III we compare these
approximations to exact solutions to demonstrate their accuracy.
Our conclusions can be found in Section IV.

\section{Growth of matter perturbations in scalar-tensor gravity beyond subhorizon scales}
\label{ScalarTensor} ST theories are widely studied as
an alternative to GR. These theories are well-motivated from
string theory, Randall-Sundrum models, as well as extended and
hyperextended inflationary models. (See e.g. \cite{faraoni} and
references therein for a review of ST theories.) The
deviations from GR predicted by these theories have been
investigated (see e.g. \cite{Will:1994fb, Damour:1995kt,
Chiba:1999wt, EspositoFarese:2004cc, Clifton:2004st, Ni:2005ej,
Turyshev:2008dr, Tsujikawa:2008uc}). ST theories have
also been used to explain the accelerating Universe, and the
cosmological consequences of these models have been widely studied
(see e.g. \cite{Boisseau:2000pr, EspositoFarese:2000ij,
Chen:1999qh, Damour:1993id, Santiago:1998ae, Damour:1998ae,
Navarro:1999ss, Perrotta:1999am, Holden:1999hm, Bartolo:1999sq,
Gaztanaga:2000vw, Capozziello:2005mj,Gannouji:2006jm,
Capozziello:2007iu, Demianski:2007mz, Barenboim:2007bu,
Jarv:2008eb, Vitagliano:2009zy,Tatsuya:2009,Song:2010}).

In this work, we focus on the growth of matter perturbations in
these theories. As shown in \cite{Boisseau:2000pr,EspositoFarese:2000ij}, if one
works in the Newtonian gauge and considers scales much smaller
than the horizon ($k\gg aH$) then the overdensity $\delta$ obeys
an equation very similar to the familiar \be
\label{growthequation} \ddot{\delta} + 2H\dot{\delta}  -4\pi G_{\rm eff}(t)
\rho_m\delta = 0\,, \ee where dots denote derivatives with
respect to cosmic time and $G_{\rm eff}(t)$ is an effective gravitational constant whose evolution is determined by the scalar field dynamics (see also equation (\ref{growthequationGeff}) below). We reconsider the growth of perturbations
in these models, manifestly retaining the scale-dependent effects,
and derive an improved version of the growth equation  which
models the evolution of $\delta$ with a greater accuracy than the
scale-independent equation. Using our improved growth equation, we
then propose a new scale-dependent parameterization for growth in
these models which is applicable under the assumption of our approximations which involve slow evolution of the scalar field.

\subsection{Scalar-tensor cosmology}
\label{ScalarTensor}
We start with the general action of a Universe described by ST gravity (in the Jordan frame) and arbitrary matter fields:
\begin{eqnarray}
S&=&\frac{1}{16\pi G_*}\int d^4x\sqrt{-g}\[F(\Phi)R\right.\\
&-&\left.Z(\Phi)g^{\mu\nu}\partial_\mu\Phi\partial_\nu\Phi-2U(\Phi)\]
+S_m\[\psi_m;g_{\mu\nu}\]\nonumber
\end{eqnarray}
where $g_{\mu\nu}$ is the metric with determinant $g$ and Ricci
scalar R. $G_*$ is the bare gravitational coupling constant
(henceforth we will set $8\pi G_{*} = 1$).  The scalar field
$\Phi$ has a potential $U(\Phi)$ and couples to gravity through
the functions $F(\Phi)$ and $Z(\Phi)$. $S_m$ denotes the action of
matter fields $\psi_m$. Henceforth we work with the parametrization
where $Z(\Phi)=1$ and $F(\Phi)$ is arbitrary.

We next consider a spatially flat Friedman-Robertson-Walker (FRW)
Universe with a background metric (in the Jordan frame): \be
ds^2=-dt^2+a^2(t)\[dr^2+r^2\(d\theta^2+\sin^2\theta d\varphi^2\)\]\,.
\ee We take the matter content of the Universe to be a perfect
fluid with energy-momentum tensor:
\be
T_{\mu\nu}\equiv\frac{2}{\sqrt{-g}}\frac{\delta S_m}{\delta g_{\mu\nu}}=\(p+\rho\)u_{\mu}u_{\nu}+p\,g_{\mu\nu}\,,
\ee
where $p$, $\rho$ and $u^\mu$ are the
pressure, energy density and four-velocity of the matter fluid
respectively. For convenience, we will henceforth consider the
matter to be pressureless and set $p=0$.

Finally, we work at linear order in perturbation theory in the
Newtonian gauge. The metric is perturbed as follows: \be
ds^2=-\(1+2\phi\)dt^2+a^2\(1-2\psi\)d\mathbf{x}^2 \ee where $\phi$
and $\psi$ are the temporal and spatial perturbations and
$\mathbf{x}\equiv\(r,\theta,\varphi\)$. The matter energy density
is perturbed as $\rho\rightarrow \rho+\delta\rho$, where $\rho$ is
the background density and $\delta\rho$ is the perturbation. For
convenience, we define the overdensity
$\delta\equiv\delta\rho/\rho$. The perturbations in the velocity
field $\delta u_{\mu}$ are conveniently expressed through the
velocity potential $v$ (defined such that $\delta
u_{\mu}=-\partial_{\mu}v$). The scalar field is perturbed as
\mbox{$\Phi\rightarrow \Phi+\delta\Phi$}.

The evolution of the background variables is governed by the zero'th order Einstein equations and the equations of conservation of energy-momentum:
\begin{align}
\label{bgd1}&3FH^2=\rho_m+\frac12\dot{\Phi}^2-3H\dot{F}+U\\
\label{bgd2}&-2F\dot{H}=\rho_m+\dot{\Phi}^2+\ddot{F}-H\dot{F}\\
\label{bgd3}&\ddot{\Phi}+3H\dot{\Phi}=3F'(\dot{H}+2H^2)-U'\\
\label{bgd4}&\dot{\rho}_m+3H\rho_m=0
\end{align}
where primes denote derivatives with respect to the scalar field $\Phi$ and dots denote derivatives with respect to the coordinate time.

The evolution of the perturbations is governed by the first order
Einstein and conservation equations. We work in Fourier space
assuming a spatial dependence of $\exp(i\mathbf{k}.\mathbf{x})$.
The scalar field fluctuation is given by

\begin{widetext}
\begin{eqnarray}
\label{fluctphi}
\delta\ddot{\Phi}+3H\delta\dot{\Phi}+\[\frac{k^2}{a^2}-3\(\dot{H}+2H^2\)F''+U''\]\delta\Phi
&=&\[\frac{k^2}{a^2}\(\phi-2\psi\)-3\(\ddot{\psi}+4H\dot{\psi}+H\dot{\phi}\)\]F'\nonumber \\&&+
\(3\dot{\psi}+\dot{\phi}\)\dot{\Phi}-2\phi U^\prime\,.
\end{eqnarray}
\end{widetext}

The matter density perturbation and velocity potential  evolve as
\ba
\label{deltadot}\dot{\delta}&=&-\frac{k^2}{a^2}v+3\dot{\psi}\\
\label{vdot}\dot{v}&=&\phi \ea

The evolution of the metric perturbations is given by the equations

\begin{widetext}
\begin{eqnarray}
\label{aniso} \psi &=&
\phi+\frac{F'\delta\Phi}{F}\\
\label{metricpert1}2F\(\dot{\psi}+H\phi\) +\dot{F}\phi &=& \rho v+\dot{\Phi}
\,\delta\Phi+\delta\dot{F}-HF'\delta\Phi\\
\label{metricpert2}-3\dot{F}\dot{\phi}-\(2\frac{k^2}{a^2}F-\dot{\Phi}^2+3H\dot{F}\)\phi
&=& \rho\(\delta+3Hv\)+U'\delta\Phi
+\(\frac{k^2}{a^2}-6H^2-3\frac{\dot{F}^2}{F^2}\)\delta
F+\dot{\Phi}\,\delta\dot{\Phi}\nonumber \\&&+3H\dot{\Phi}\,\delta\Phi+3\frac{\dot{F}}{F}\delta{\dot{F}}
\end{eqnarray}
\end{widetext}

\subsection{Growth of matter perturbations}

Note that Eqns.(\ref{deltadot},\ref{vdot}) can be combined to
obtain an exact second-order differential equation for $\delta$:
\be
\label{deltaexact}\ddot{\delta}+2H\dot{\delta}+\frac{k^2}{a^2}\phi=3\ddot{\psi}+6H\dot{\psi}
\ee

We now proceed to obtain an approximate equation for the growth of
perturbations of subhorizon modes. However, instead of considering
only terms which are proportional to $k^2/a^2$, as is usually done
(see e.g. \cite{Copeland,EspositoFarese:2000ij}), we also retain
terms proportional to $H^2$. This makes the analysis more accurate
for sub-Hubble modes close to the Horizon size. In addition, we
also assume matter-domination, during which the metric potentials
are frozen. Thus we ignore time derivatives of metric perturbations. We also ignore time derivatives of the field. Even though this approximation is not always applicable, in practice we have found that our analytical scale dependent approximation for the growth of perturbations turns out to always be significantly better than the usual scale independent sub-Hubble approximation. The precise accuracy level however depends on the degree of validity of the above approximation which may vary depending on the details of the field dynamics.

Defining 
\be
\xi(a,k) \equiv 3a^2 H(a)^2/k^2\simeq \frac{3 H_0^2 \Omega_{0m}}{a k^2}
\ee
(where the last equation is valid in the matter era) \eqn{fluctphi} reduces to:

\begin{align}
&\delta\Phi=\nonumber\\
&-\phi\(\frac{FF'+2\xi FU'/3H^2}{F+2F'^2+\xi\(FU''/3H^2-F''U/2H^2-F F''/2\)}\)\,.\nonumber\\
\end{align}
Plugging into \eqn{metricpert2} we obtain \be
\frac{k^2}{a^2}\phi=-\frac{1}{2F}\,\rho\(\delta+3Hv\)g\(F,U,\xi\)\,,
\ee where
\begin{widetext}
\begin{equation*}
g\(F,U,\xi\)\equiv
\frac{2F+4F'^2+\xi\(2FU''/3H^2-FF''-F''U/H^2\)}{2F+3F'^2+\xi
\(2F'^2-FF''+2FU''/3H^2-U'F'/H^2-F''U/H^2\)+\xi^2\(4F'U'/3H^2-2U'^2/9H^4\)}\,.
\end{equation*}
\end{widetext}
Using the above, (and eliminating $v$ from \eqn{metricpert1}
ignoring time derivatives), we obtain the following form for the
Poisson equation: \be \frac{k^2}{a^2}\phi=-
\frac{1}{2F}\rho\delta\(\frac{g\(F,U,\xi\)}{1+\xi
g\(F,U,\xi\)h\(F,U,\xi\)}\)\,, \ee where
\begin{widetext}
\begin{equation*}
h\(F,U,\xi\)\equiv
\frac{2F+3F'^2+\xi\(2FU''/3H^2-2F'U'/3H^2-F''U/H^2-FF''\)}{2F+4F'^2+\xi\(2FU''/3H^2-F''U/H^2-FF''\)}\,.
\end{equation*}
\end{widetext}

This leads to our ``improved'' growth equation for the evolution
of perturbations, accurate for large subhorizon scales: \be
\label{improvedgrowthequation}
\ddot{\delta}+2H\dot{\delta}-\frac{1}{2F}\[\frac{g\(F,U,\xi\)}{1+\xi
g\(F,U,\xi\)h\(F,U,\xi\)}\]\rho\delta=0\,. \ee 
In the case of general relativity this reduces to the form derived in Ref. \cite{DentDutta:2008,DentDuttaPerivolaropoulos:2009}
\be {\ddot \delta} + 2 H {\dot \delta} - \frac{4\pi G \rho_m \delta}{1+\xi(a,k)} =0 \label{greq-scdepd} \ee
Note that for scales
much smaller than the horizon (\mbox{$k\gg aH$}, or equivalently
$\xi\rightarrow 0$) equation (\ref{improvedgrowthequation}) reduces to the well-known form (see e.g.
Eqn.(5.13) of \cite{EspositoFarese:2000ij}): \be
\label{growthequationGeff}
\ddot{\delta}+2H\dot{\delta}-\frac{1}{2F}\[\frac{2F+4F'^2}{2F+3F'^2}\]\rho\delta=0\,.
\ee

Equation (\ref{greq-scdepd}) may be expressed in terms of the growth factor $f=\frac{d\ln \delta}{d\ln a}$ in the form
\be f' + f^2 + \(2-\frac{3}{2} \omms(a)\)f=\frac{3}{2}\frac{\omms(a)}{1+\xi(a,k)} \label{greqlna-scdepf} \ee
where $'\equiv \frac{d}{d\ln a}$, and we have assumed \lcdm for $H(a)$.  

For sub-Hubble scales $\xi(k,a)\rightarrow 0$ and the solution of equation (\ref{greqlna-scdepf}) is well approximated by (\ref{stdpar}) with $\gamma=\frac{6}{11}$. For larger scales a perturbative approach \cite{DentDuttaPerivolaropoulos:2009} may be used to derive the scale dependent growth rate $f(a,k)$ of matter perturbations in GR as: \be \label{newpargr}
f(a,k)=\Omega_{m}\(a\)^\gamma \(1+\frac{3 H_0^2 \Omega_{0m}K}{a k^2}\)^{-1}\,, \ee with $K=0$ for the scale-independent growth
function and \mbox{$K=1$} for the GR scale-dependent
approximation.
Following the same reasoning as in Ref. \cite{DentDuttaPerivolaropoulos:2009}, the following parametrization may be derived for the growth function $f(a,k)$ in ST cosmologies: \be \label{newparst} f(a,k)=\Omega_m(a)^\gamma
\frac{1}{F}\(\frac{g\(F,U,\xi\)}{1+\xi g\(F,U,\xi\)h\(F,U,\xi\)}\).
\ee
Even though this parametrization may be derived as an approximate solution by a perturbative expansion to all orders in $\xi$ as in Ref. \cite{DentDuttaPerivolaropoulos:2009}, its validity for $\gamma\simeq 0.55$ (the exact value is obtained by fitting to the exact numerical solution) will also be tested in the next section. The scale dependence of $\gamma$ has also been recently discussed in \cite{bagram,weller}.
\begin{figure}[htbp]
\epsfig{file=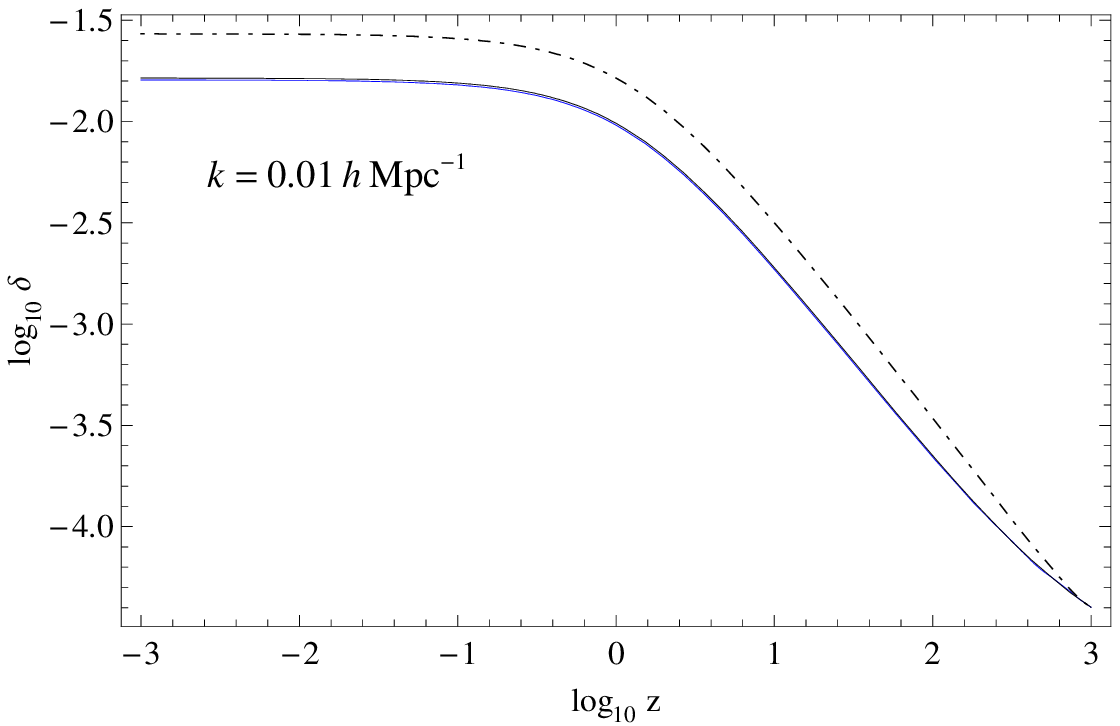,width=8cm}
\epsfig{file=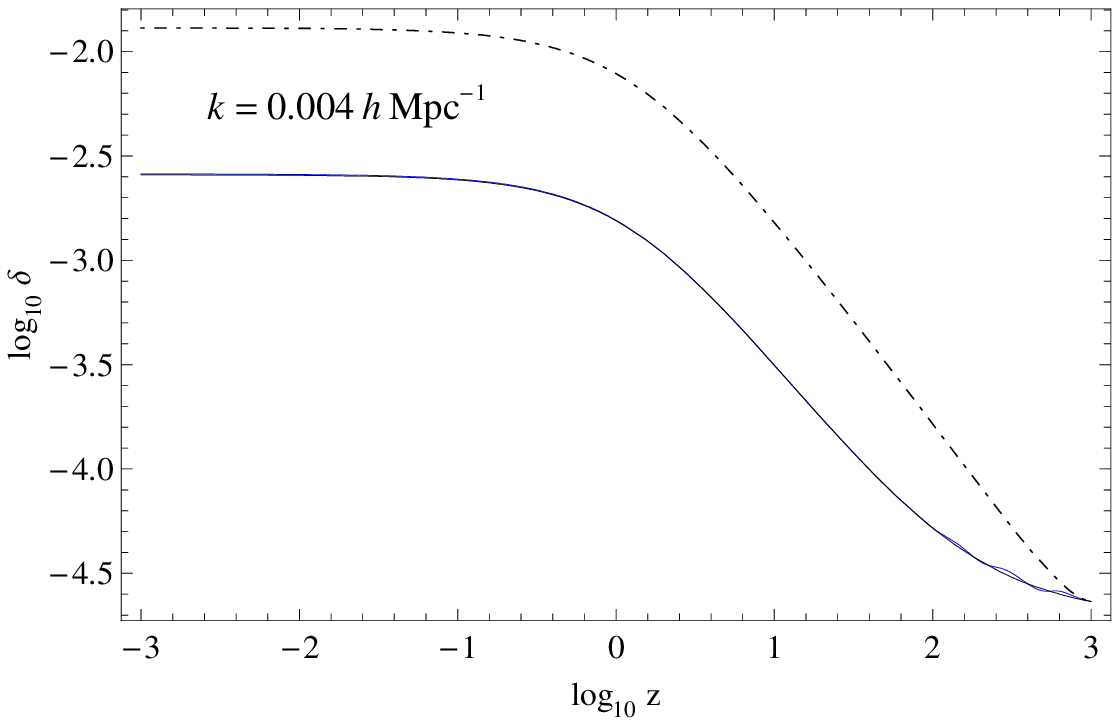,width=8cm}
\epsfig{file=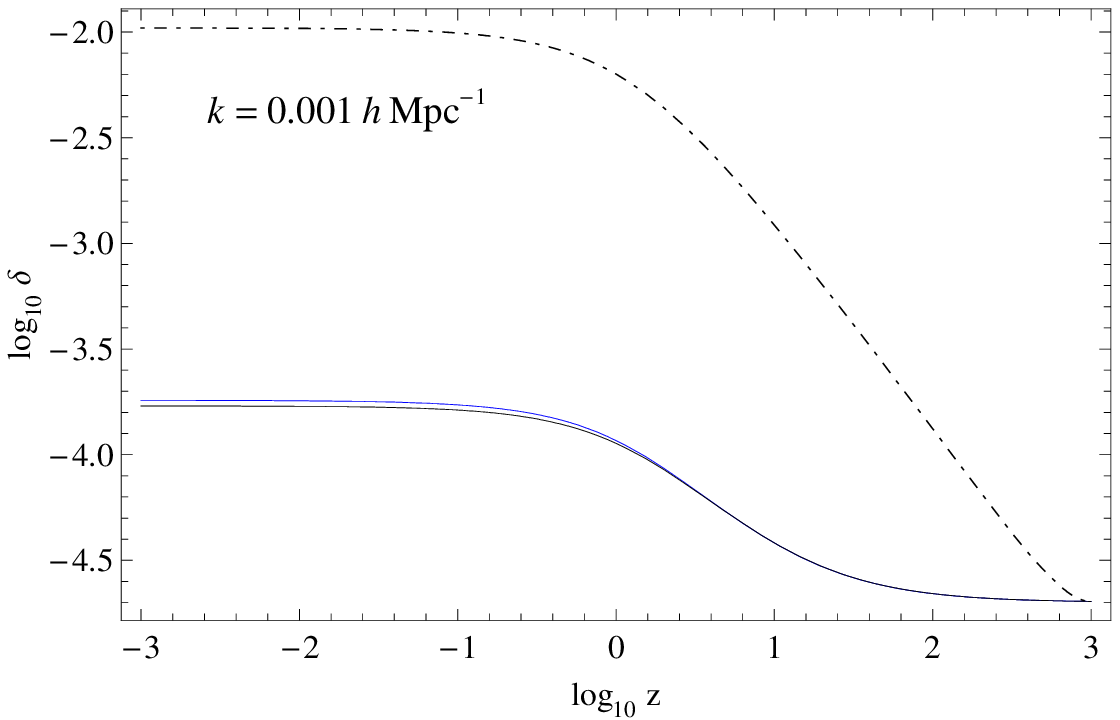,width=8cm} \caption{For three choices
of scale, plot of the exact $\delta_m(k,a)$ (blue solid line),
along with two approximations: (i) the solution of the
scale-dependent ST growth equation
\eqref{improvedgrowthequation} (solid line) and
(ii) the solution of the scale-independent ST sub-Hubble growth equation
(dot-dashed line). The apparent scale dependence of the present time value of $\delta$ in the sub-Hubble approximation is due to the scale dependent initial conditions considered needed to secure a scale independent value of the initial metric perturbations \cite{Sanchez:2010ng}}\label{figure1}
\end{figure}

\begin{figure}[htbp]
\epsfig{file=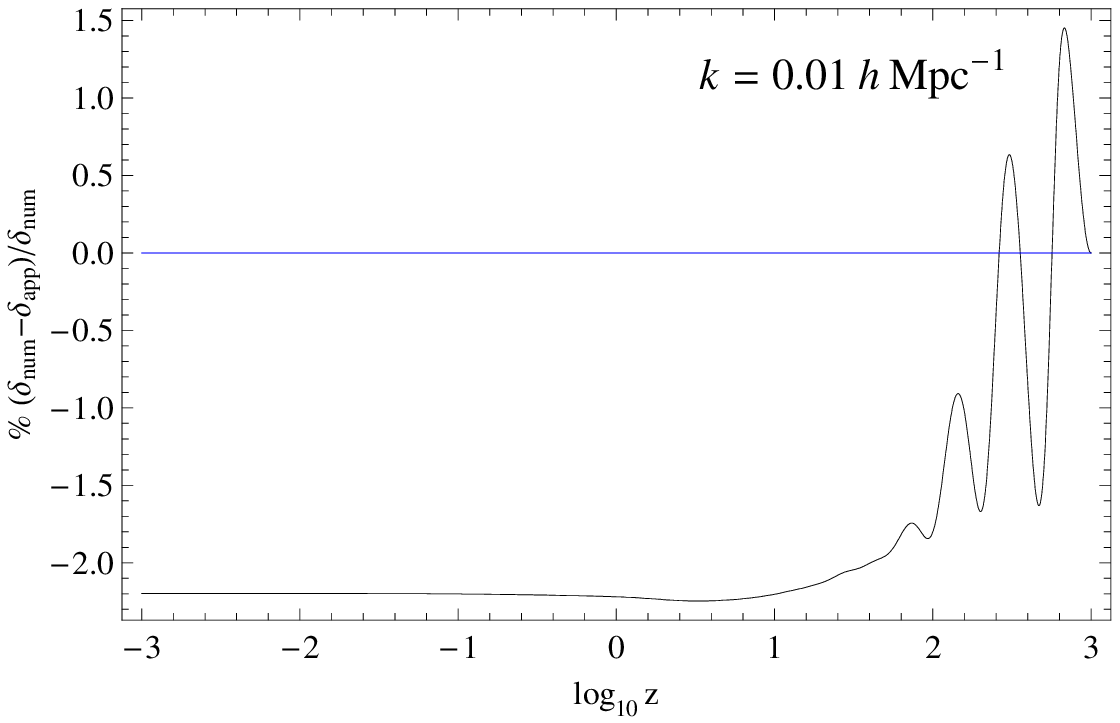,width=8cm}
\epsfig{file=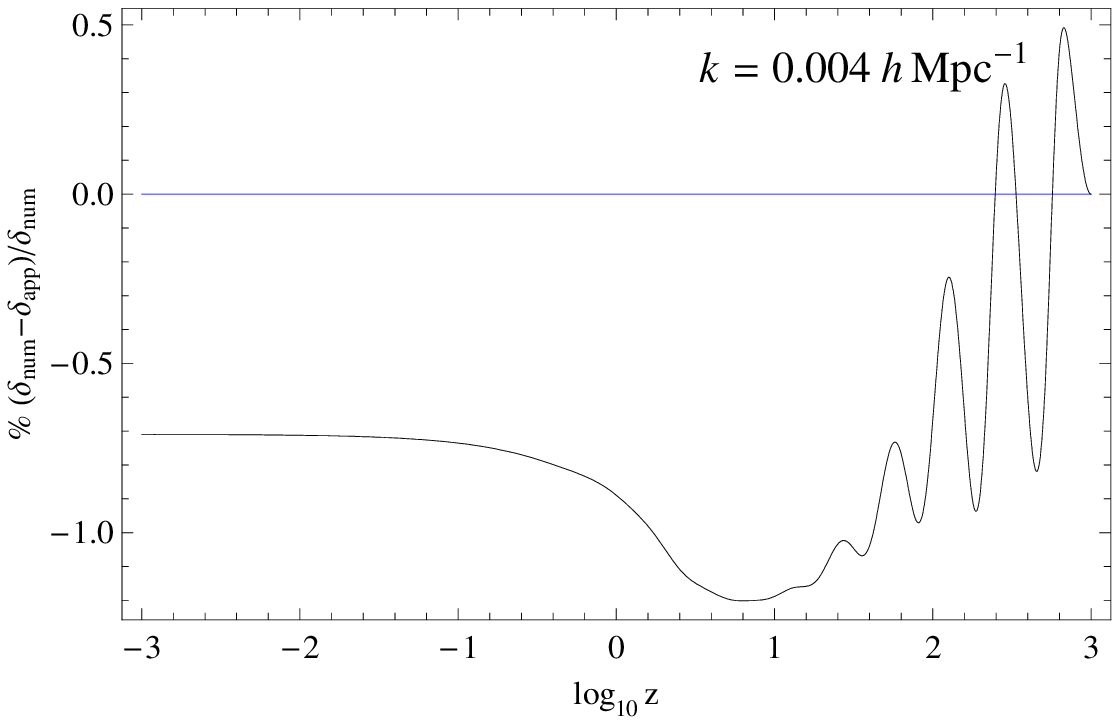,width=8cm}
\epsfig{file=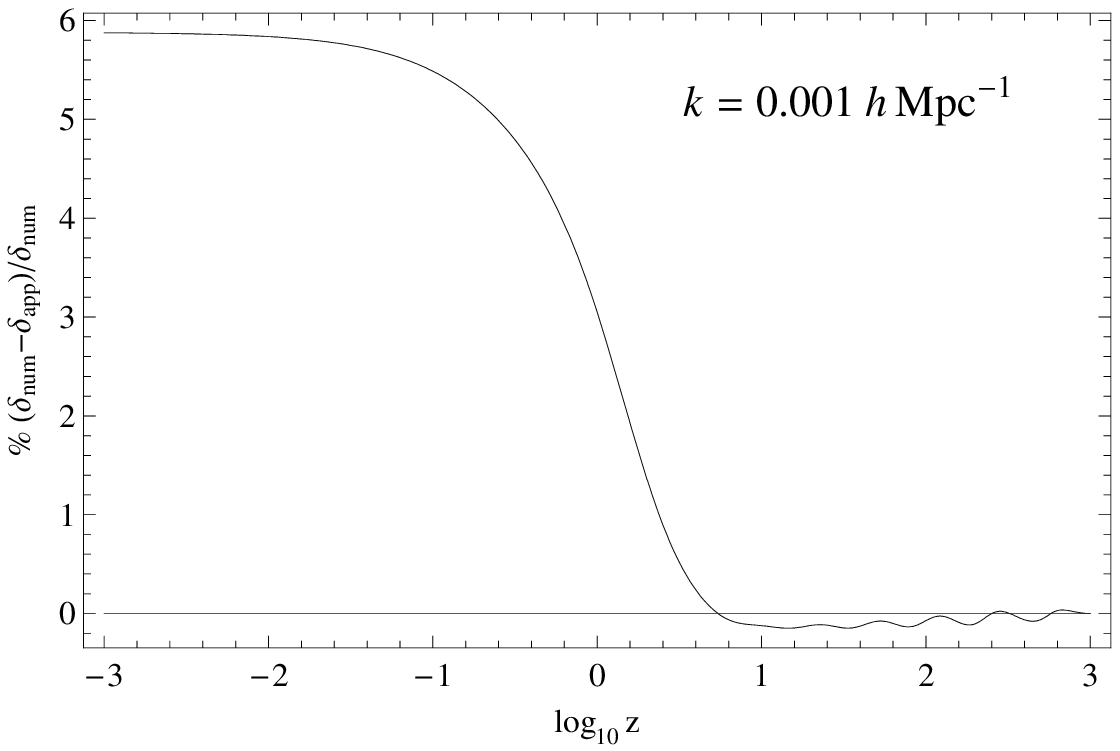,width=8cm} \caption{Plots of the $\%$
difference between the exact numerical solution for $\delta_m$
(blue line) and the scale-dependent ST approximation (solid
line). The
$\%$ difference corresponding to the subhorizon ST approximation
is not included since this is much bigger than the corresponding
to the scale-dependent approximations. The values used for the
plot are $\lambda_f=5$ and $\lambda=10$.
}\label{figure2}
\end{figure}

\begin{figure}[htbp]
\epsfig{file=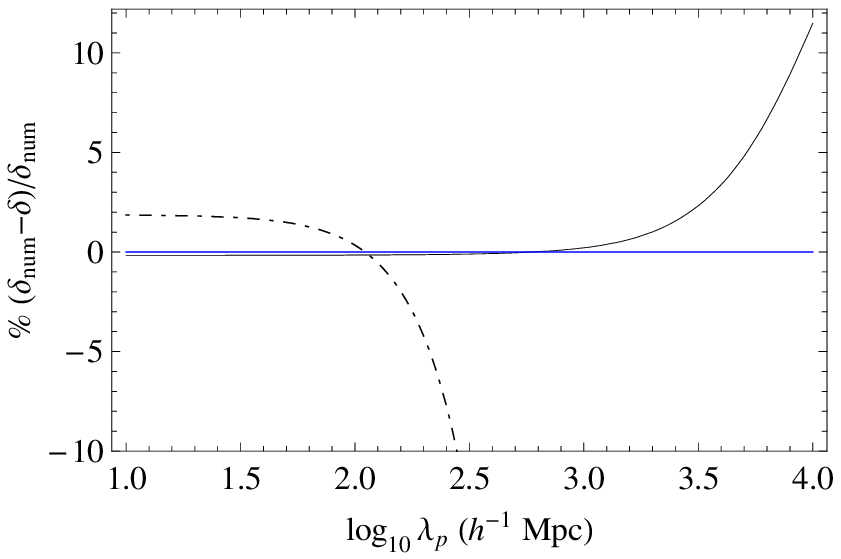,width=8cm}
\epsfig{file=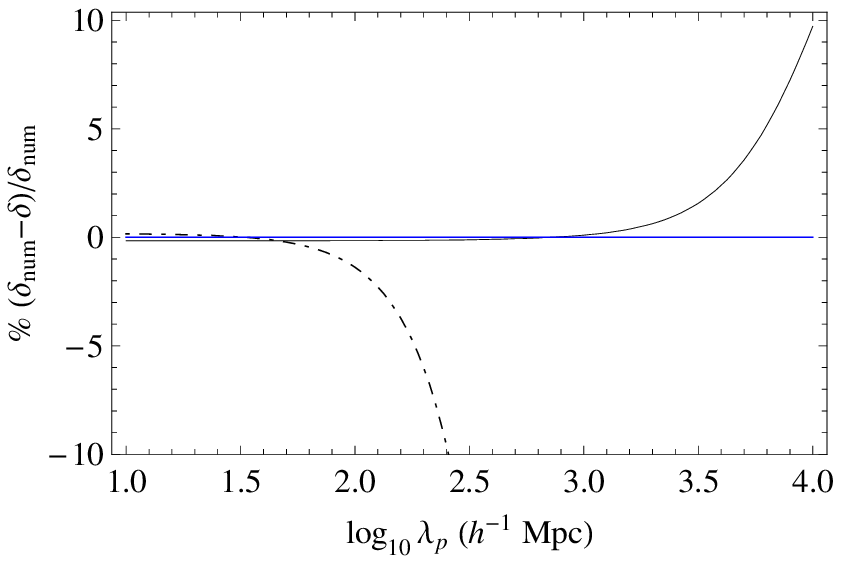,width=8cm} \caption{Plots of the $\%$
difference between numerical solution for $\delta$ (in blue) and
(i) the scale-dependent ST approximation (solid line) and (ii) the
subhorizon ST approximation (dot dashed line) as a function of the
perturbation scale $\lambda_p\equiv\frac{2\pi}{k}$. We set $\lambda_f=5$ in both plots and
$\lambda=10$ and $\lambda=1$ for the top and bottom panels,
respectively.}\label{figure3}
\end{figure}
\section{Comparison to exact results}
\label{comparison}

We now compare the evolution predicted by
\eqn{improvedgrowthequation} and \eqn{newparst} to results from the
exact evolution.

To solve the system numerically, we choose the following
functional forms \cite{Sanchez:2010ng}. We take \be
F(\Phi)=1-\lambda_f\Phi^2\,\,;\,\,Z(\Phi)=1\,\,;\,\,U(\Phi)=1+\exp[-\lambda\Phi] \label{potentials}
\ee Then we solve numerically the background equations given by
\eqn{bgd1}-\eqn{bgd4} (for details see \cite{Sanchez:2010ng}).

To evolve the perturbations, we solve equations \eqn{metricpert1},
\eqn{metricpert2} and \eqn{deltaexact}. These are solved in the Newtonian gauge using the background. The initial time is taken
to be in the matter dominated epoch, at $z_i=1000$, and we set
$\dot{\Phi}_i\simeq0$.

In our numeric solution the initial value
$\Phi_i$ is chosen so that at early times the deviation in the
expansion rate from the $\Lambda$CDM one is not larger
than around $5\%$. This deviation is roughly given by
$\frac{H_{ST}-H_{\textrm{$\Lambda$CDM}}}{H_{\textrm{$\Lambda$CDM}}}\sim
F(\Phi_i)^{-1/2}-1$, which then sets an upper bound $\Phi_{i,{\rm
max}}$ for $\Phi$. Although $\Phi_i/\Phi_{i,{\rm max}}\sim1$
results in an expansion rate consistent with the $\Lambda$CDM one,
we find that such a choice gives rise to deviations between the
analytic approach in Eq.~(\ref{improvedgrowthequation}) and the
numeric solution for $\delta_m$. This is because larger $\Phi$ implies larger effective potential energy of the field, and hence larger $\dot{\Phi}$ when the field oscillates due to its coupling to curvature and therefore to matter. Above a certain threshold, $\dot{\Phi}$ grows large enough so that one cannot keep neglecting it in order to arrive to Eq.~(\ref{improvedgrowthequation}). In what follows we consider $\Phi_i/\Phi_{i,{\rm max}}\lesssim0.30$. For this initial condition we secure relatively small deviation from GR and from the $\Lambda$CDM expansion rate (consistent with observations) and small field time derivative (consistent with our approximation). We thus find that the solution to Eq.~(\ref{improvedgrowthequation}) deviates from the numeric solution for $\delta_m$ by less than $1\%$ on subhorizon scales.

We first compare the exact numerical
solution of the ST perturbation system (\ref{metricpert1}), (\ref{metricpert2}) (potentials given by (\ref{potentials})) with the solution of the approximate linear growth equation  assuming the following approximations:
\begin{itemize}
\item Scale Dependent ST growth equation (\ref{improvedgrowthequation}).
\item Scale independent ST sub-Hubble approximation (\ref{growthequationGeff}) \end{itemize}
This comparison is shown in detail in Figures \ref{figure1} - \ref{figure3}. In Figure 1 we show the evolution of the matter density perturbation $\delta(z)$ in the above three cases (two approximations and the exact solution). For intermediate scales ($k=0.01 h^{-1}$Mpc, upper panel of Figure \ref{figure1}) the scale dependent approximation of equation (\ref{improvedgrowthequation}) (black continuous line for ST) fit well the exact numerical result (blue continuous line) while the scale independent sub-Hubble approximation (dot-dashed line) is significantly less accurate. On larger scales, the sub-Hubble approximation becomes even less accurate (lower panel of Figure \ref{figure1}) while the scale dependent ST approximation of eq. (\ref{improvedgrowthequation}) remains a good approximation to the exact numerical solution within a few percent (Fig. \ref{figure2}).  This is more clearly demonstrated in Figure \ref{figure3} where we show the scale dependence of the deviation of the two approximate present values of $\delta$ from the corresponding exact value of $\delta$. Clearly  the scale dependent approximation is in good agreement with the exact solution on physical scales up to $\lambda_p=10^3 h^{-1}$Mpc ($\lambda_p\equiv\frac{2\pi}{k}$). On the other hand, the sub-Hubble scale independent approximation (dot-dashed line in Figure \ref{figure3}) starts deviating significantly from the exact solution already on scales larger than about $10^2 h^{-1}$Mpc. On scales significantly larger than $10^3 h^{-1}$Mpc, both approximations break down as shown in Figure \ref{figure3}.

\begin{figure}
\epsfig{file=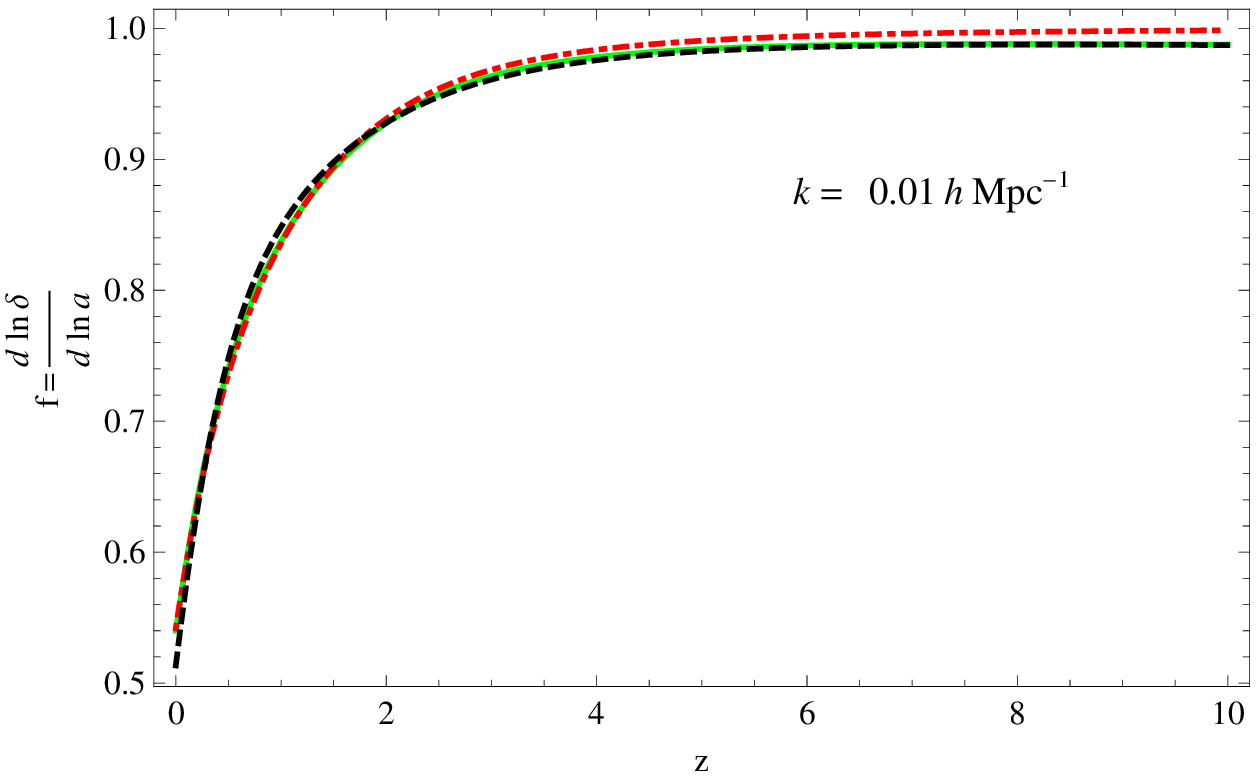,width=8cm}
\epsfig{file=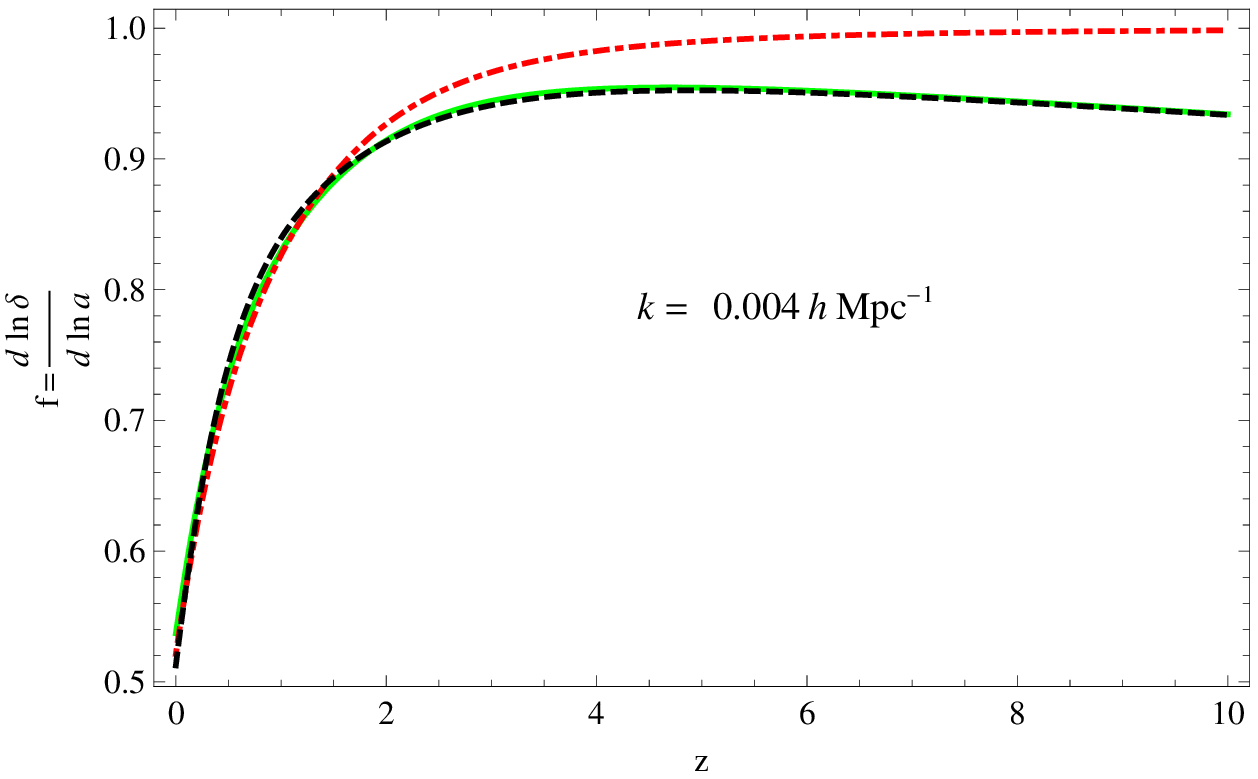,width=8cm}
\epsfig{file=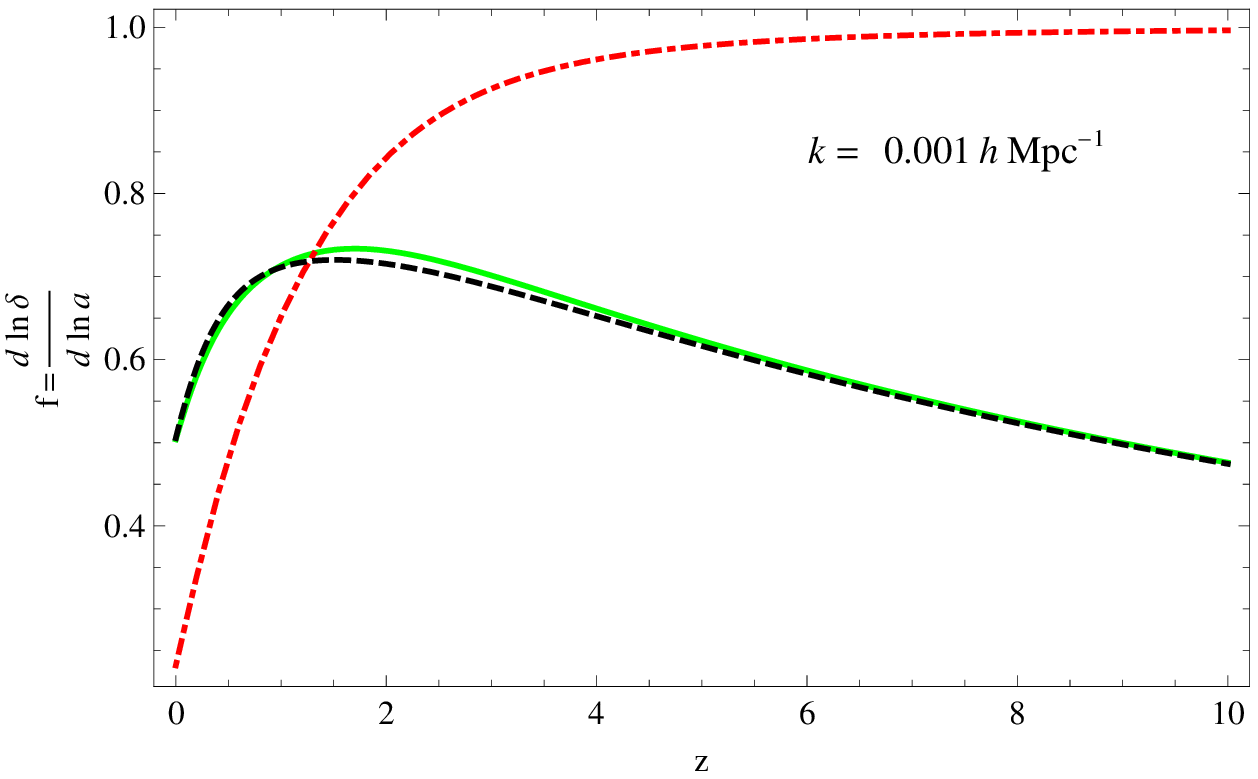,width=8cm} \caption{For three choices
of scale, plot of the exact growth function $f(k,a)$ (green, solid
line), along with two approximations: from the scale-dependent
ST growth equation \eqref{newparst}
(black, dashed line)
and from the scale-independent GR sub-Hubble growth equation \eqref{newpargr} with $K=0$ (red, dot-dashed
line).}\label{figure4}
\end{figure}

The approximate scale dependent growth rate $f(a,k)$ obtained in equation (\ref{newparst}) is compared with the corresponding exact result in Figure \ref{figure4} where we show a superposition of the growth factor $f(a,k)$ corresponding to the approximations (sub-Hubble scale independent ST and scale dependent ST) and to the exact solution. The parameter values used in Figure \ref{figure4} are $\lambda=10$, $\lambda_f =1$ which provide a background evolution similar to the best fit $\Lambda CDM$ model (the difference for the expansion rate is always less than $5\%$). Clearly, the scale independent growth factor deviates significantly from the exact result while the scale dependent approximation remains very close to it even on large scales ($k=0.001 h\,$Mpc$^{-1}$) and redshifts. Thus even though on relatively small scales ($k>0.01 h\,$Mpc$^{-1}$) or low redshifts ($z<2$) the accuracy of the two approximations is good and similar, on larger scales and redshifts the accuracy of the scale dependent approximation is significantly better than the scale independent approximation which fails to approximate the exact numerical solution.

\section{Connection to the Synchronous Gauge}
\label{comparison}

The above calculations have been made in the conformal Newtonian gauge. In this section we briefly discuss the gauge dependence of the density perturbation $\delta$. Let us consider the gauge invariant quantity \be \delta_m\equiv \frac{\delta \rho}{\rho + p} + 3 H v \label{ginvpert} \ee where $v$ is the velocity potential. This quantity may be evaluated in the synchronous and in the Newtonian gauges leading to \be \delta_m^{(S)} = (\frac{\delta \rho}{\rho + p})^{(S)}=\delta_m^{(N)}=(\frac{\delta \rho}{\rho + p})^{(N)} + (3 H v)^{(N)} \label{giconx} \ee where the superscripts $^{(S)}$ and $^{(N)}$ imply evaluation in the Synchronous and in the Newtonian gauge respectively. Assuming matter domination ($p=0$) and dropping the index $^{(N)}$ for the Newtonian gauge we find \be \delta^{(S)}=\delta + 3 H v \label{con2} \ee

The evolution equation for matter perturbations in the synchronous gauge is
\be {\ddot \delta^{(S)}} + 2 H {\dot \delta^{(S)}} + 4 \pi G \delta^{(S)} =0 \label{synchgr} \ee which is scale independent. The corresponding equation in the Newtonian gauge is eq. (\ref{deltaexact}). When properly related initial conditions are used in the solutions of eqs. (\ref{deltaexact}) and (\ref{synchgr}), their solutions should satisfy eq. (\ref{con2}).

Consider for example initial conditions of the form \be \delta \Phi_i = {\dot {\delta \Phi_i}}=\dot{\psi}_i =0, \;\;\;\;  \psi_i \simeq -10^{-5}= {\rm const} \label{initcondex} \ee

\begin{figure}
\epsfig{file=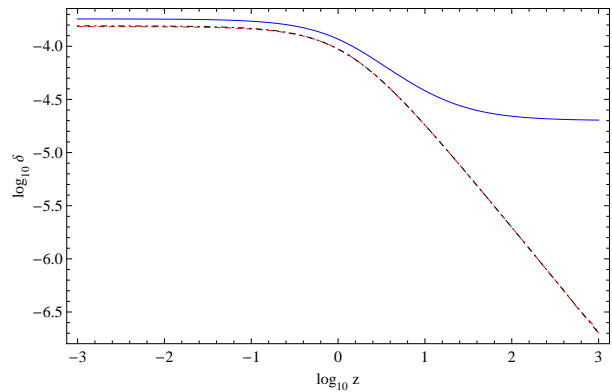,width=8cm}
\label{comparison}
 \caption{For a large scale ($k=0.001 h Mpc^{-1}$) we show the evolution (with parameter values $\lambda_f=5$, $\lambda=10$) of $\delta(a)$ (blue continuous line), $\delta^{(S)}(a)$ (black dot-dashed line) and $\delta(a) + 3 H v$ (red dashed line which coincides with the line of $\delta^{(S)}(a)$ as anticipated).}\label{figure5}
\end{figure}

It is straightforward to show (see Appendix) that the above initial conditions in the matter era transform to the following initial conditions for the matter perturbations in each gauge:
\ba
\label{NewtIC1}
\delta_i^{(S)} & = & -\(\frac{2 k^2}{3 a_i^2 H_i^2}\)\psi_i \\
\label{NewtIC2}
{\dot \delta}_i^{(S)} & = & -\(\frac{2 k^2}{3 a_i^2 H_i^2}\)H_i \psi_i
\ea
for the synchronous gauge and
\ba
\label{SynchIC1}
\delta_i & = & -\(\frac{2 k^2}{3 a_i^2 H_i^2}+2\)\psi_i \\
\label{SynchIC2}
{\dot \delta}_i & = & -\(\frac{2 k^2}{3 a_i^2 H_i^2}\)H_i \psi_i
\ea
for the Newtonian gauge. Using these initial conditions we solve the perturbation equation in each gauge (eqs. (\ref{deltaexact}) and (\ref{synchgr})) and verify the validity of eq. (\ref{con2}). This is an additional verification of the validity of our analysis and demonstrates that the difference of $3 H v$ between the matter density perturbations in the two gauges can be significant. It is demonstrated in Fig. \ref{figure5} where we consider a large scale ($k=0.001 \; h \;  Mpc^{-1}$) and show the evolution of $\delta(a)$ (Newtonian gauge, blue continuous line), $\delta^{(S)}(a)$ (Synchronous gauge, black dot-dashed line) and $\delta(a) + 3 H v$ (red dashed line which coincides with the line of $\delta^{(S)}(a)$ as anticipated).

\section{Conclusions}
\label{conclusions}
We have derived and tested numerically a simplified ordinary differential equation whose solution describes fairly accurately the growth of linear cosmological perturbations in ST cosmologies up to Hubble scales (beyond $10^3 h^{-1}$Mpc). A corresponding analytic form of the scale dependent growth rate $f(a,k)$ was also presented and tested numerically in comparison with the exact linear result. This is a significant improvement over the previously known sub-Hubble which breaks down on scales larger than about $10^2 h^{-1}$Mpc.

We have thus demonstrated that in the context of the Newtonian gauge, the comparison of cosmological large scale structure data with corresponding theoretical predictions in ST based cosmologies on scales larger than a few hundred Mpc should not be based on the sub-Hubble scale independent approximation. Instead it requires either use of the full linear numerical solution of the cosmological perturbation equations in each theory or the imposed scale dependent approximation presented in the present study. Thus, previous work using the scale independent sub-Hubble approximation to make predictions for the growth of perturbations in ST theories is reliable on scales up about $200h^{-1}Mpc$. However, such results should be used with care on larger scales. For example, on scales of a few hundred $Mpc$ the error induced in the growth rate by not using the scale dependent effects is about $10\%$ on redshifts $z>3$ while on scales larger than $1Gpc$ the corresponding error exceeds $50\%$ even at low redshifts $z\simeq 1$.

We have also shown that our results are consistent with a corresponding calculation in the synchronous gauge and verified that the quantity $\frac{\delta \rho}{\rho}$ is gauge dependent and there can be a significant difference between its forms in different gauges on large scales.

\section*{Acknowledgements}
JCBS and LP are supported by the European Research and
Training Network MRTPN-CT-2006 035863-1 (UniverseNet).

\appendix
\section{Growth in synchronous and conformal Newtonian gauges}

In this appendix we discuss the appropriate (scale-dependent) initial conditions that must be used in the Newtonian and synchronous gauges in order to verify \eqn{con2}, i.e., equations (\ref{NewtIC1})-(\ref{SynchIC2}).

Starting with Eq.(\ref{metricpert2}), and using the initial conditions

\begin{equation}
\label{IC}
\delta \Phi_i = {\dot {\delta \Phi_i}}=\dot{\psi}_i =0, \;\;\;\;  \psi_i \simeq -10^{-5}\neq 0
\end{equation}
(a subscript \emph{i} will always refer to the value of a quantity at the initial time) we obtain the following relation for $\delta_i$

\be
\label{five}
\delta_{i}=\frac{1}{\rho_{mi}}\(2\frac{k^2}{a^2}F_i-\dot{\Phi}_{i}^2+3H_i\dot{F}_i\)\psi_i+3H_i\, v_i\, \rho_{mi}
\ee

Using Eq.(\ref{metricpert1}) with the above initial conditions (\ref{IC}) we arrive at the relation

\be
\label{six}
3H_iv_i=\frac{1}{\rho_{mi}}\(6H_i^2F_i+3H_iF^{\prime}_i\dot{\Phi}_i\)\psi_i
\ee

Next combining Eqns. (\ref{five}) and (\ref{six}), and ignoring time derivatives, we can then recover the initial condition for $\delta$:
\be
\delta_{i}\simeq-\left(2\frac{k^2}{a^2}+6H_i^2\)\frac{F_i\,\psi_i}{\rho_{mi}}
\ee

From \eqn{six}, using Eqs.(\ref{deltadot}), (\ref{vdot}), and (\ref{aniso}), as well as the initial conditions (\ref{IC}), we find

\be
\label{eight}
\dot{\delta}_i=-\frac{k^2}{a^2}\frac{1}{\rho_{mi}}\left(2H_iF_i+F'_I\dot{\Phi}_i\right)\psi_i
\ee

Also from Eqs.(\ref{bgd1}) and (\ref{bgd2}), ignoring time derivatives, we have the approximate relationship

\be
\label{nice}
\dot{H}_i\simeq\frac{1}{2F_i}\rho_{mi}\simeq-\frac32 H_i^2
\ee

Lastly, ignoring time derivatives in Eqns. (\ref{six}) and (\ref{eight}), and using \eqn{nice}, we obtain equations (\ref{NewtIC1}) and (\ref{NewtIC2}).

For the synchronous gauge, we start with (\eqn{con2}), use equations (\ref{five}) and (\ref{nice}), and ignore time derivatives to obtain \eqn{SynchIC1}.  Then from Eqn. (\ref{deltadot}), using equations (\ref{IC}), (\ref{six}), (\ref{eight}) and (\ref{nice}), and ignoring time derivatives, we obtain \eqn{SynchIC2}.



\begin{thebibliography}{99}


\bibitem{union08}
  M.~Kowalski {\it et al.},
  Astrophys.\ J.\  {\bf 686}, 749 (2008)
  [arXiv:0804.4142 [astro-ph]].

\bibitem{perivol}
  L.~Perivolaropoulos and A.~Shafieloo,
  arXiv:0811.2802 [astro-ph].

\bibitem{hicken}
M.~Hicken {\it et al.},
  arXiv:0901.4804 [astro-ph.CO].



\bibitem{Komatsu}
  E.~Komatsu {\it et al.}  [WMAP Collaboration],
  Astrophys.\ J.\ Suppl.\  {\bf 180}, 330 (2009)
  [arXiv:0803.0547 [astro-ph]].




\bibitem{bao}
D.~J.~Eisenstein {\it et al.}  [SDSS Collaboration],
  Astrophys.\ J.\  {\bf 633}, 560 (2005)
  [arXiv:astro-ph/0501171].

\bibitem{percival}
W.~J.~Percival, S.~Cole, D.~J.~Eisenstein, R.~C.~Nichol, J.~A.~Peacock, A.~C.~Pope and A.~S.~Szalay,
  Mon.\ Not.\ Roy.\ Astron.\ Soc.\  {\bf 381}, 1053 (2007)
  [arXiv:0705.3323 [astro-ph]].



\bibitem{Samushia2007}
  L.~Samushia, G.~Chen and B.~Ratra,
  arXiv:0706.1963 [astro-ph].

\bibitem{Ettori}
  S.~Ettori {\it et al.},
  arXiv:0904.2740 [astro-ph.CO].


\bibitem{Wang}
  Y.~Wang,
  Phys.\ Rev.\  D {\bf 78}, 123532 (2008)
  [arXiv:0809.0657 [astro-ph]].

\bibitem{Samushia2009}
  L.~Samushia and B.~Ratra,
  arXiv:0905.3836 [astro-ph.CO].

\bibitem{RatraPeebles}
  B.~Ratra and P.~J.~E.~Peebles,
  Phys.\ Rev.\  D {\bf 37}, 3406 (1988).

\bibitem{Caldwell:1997ii}
  R.~R.~Caldwell, R.~Dave and P.~J.~Steinhardt,
  Phys.\ Rev.\ Lett.\  {\bf 80}, 1582 (1998)
  [arXiv:astro-ph/9708069].

\bibitem{SteinhardtWangZlatev}
  P.~J.~Steinhardt, L.~M.~Wang and I.~Zlatev,
  Phys.\ Rev.\  D {\bf 59}, 123504 (1999).

\bibitem{WangSteinhardt:1998}
 L.~Wang and P.~J.~Steinhardt,
 Ap.\ J. {\bf{508}} (1998) 483




\bibitem{dutta}
  S.~Dutta and R.~J.~Scherrer,
  Phys.\ Rev.\  D {\bf 78}, 123525 (2008)
  [arXiv:0809.4441 [astro-ph]];
  S.~Dutta and R.~J.~Scherrer,
  Phys.\ Rev.\  D {\bf 78}, 083512 (2008)
  [arXiv:0805.0763 [astro-ph]];
  S.~Dutta and R.~J.~Scherrer,
  Phys.\ Lett.\  B {\bf 676}, 12 (2009)
  [arXiv:0902.1004 [astro-ph.CO]];
  S.~Dutta, S.~D.~H.~Hsu, D.~Reeb and R.~J.~Scherrer,
  Phys.\ Rev.\  D {\bf 79}, 103504 (2009)
  [arXiv:0902.4699 [astro-ph.CO]];
S.~Dutta, E.~N.~Saridakis and R.~J.~Scherrer,
  arXiv:0903.3412 [astro-ph.CO];
    T.~Chiba, S.~Dutta and R.~J.~Scherrer,
  Phys.\ Rev.\  D {\bf 80}, 043517 (2009)
  [arXiv:0906.0628 [astro-ph.CO]];
    D.~C.~Dai, S.~Dutta and D.~Stojkovic,
  Phys.\ Rev.\  D {\bf 80}, 063522 (2009)
  [arXiv:0909.0024 [astro-ph.CO]].


\bibitem{Parker:1999td}
  L.~Parker and A.~Raval,
  Phys.\ Rev.\  D {\bf 60}, 063512 (1999)
  [Erratum-ibid.\  D {\bf 67}, 029901 (2003)]
  [arXiv:gr-qc/9905031].

\bibitem{Boisseau:2000pr}
  B.~Boisseau, G.~Esposito-Farese, D.~Polarski and A.~A.~Starobinsky,
  Phys.\ Rev.\ Lett.\  {\bf 85}, 2236 (2000)
  [arXiv:gr-qc/0001066].

\bibitem{EspositoFarese:2000ij}
  G.~Esposito-Farese and D.~Polarski,
  Phys.\ Rev.\  D {\bf 63} (2001) 063504
  [arXiv:gr-qc/0009034].

\bibitem{Dvali:2000hr}
  G.~R.~Dvali, G.~Gabadadze and M.~Porrati,
  Phys.\ Lett.\  B {\bf 485}, 208 (2000)
  [arXiv:hep-th/0005016].

\bibitem{Freese:2002sq}
  K.~Freese and M.~Lewis,
   ``Cardassian Expansion: a Model in which the Universe is Flat, Matter
  Phys.\ Lett.\  B {\bf 540}, 1 (2002)
  [arXiv:astro-ph/0201229].

 \bibitem{Carroll:2004de}
  S.~M.~Carroll, A.~De Felice, V.~Duvvuri, D.~A.~Easson, M.~Trodden and M.~S.~Turner,
  Phys.\ Rev.\  D {\bf 71}, 063513 (2005)
  [arXiv:astro-ph/0410031].


\bibitem{OCallaghan:2009bu}
  E.~O'Callaghan, R.~Gregory and A.~Pourtsidou,
  JCAP {\bf 0909}, 020 (2009)
  [arXiv:0904.4182 [astro-ph.CO]].
  
\bibitem{HL}
  P.~Horava,
  [arXiv:0811.2217];
  P.~Horava,
  JHEP {\bf 0903}, 020 (2009)
  [arXiv:0812.4287 [hep-th]];
  P.~Horava,
  Phys.\ Rev.\  D {\bf 79}, 084008 (2009)
  [arXiv:0901.3775 [hep-th]];
  P.~Ho\v rava,
  arXiv:0902.3657 [hep-th];
  G.~Calcagni,
  arXiv:0904.0829 [hep-th];
  E.~Kiritsis and G.~Kofinas,
  arXiv:0904.1334 [hep-th];
  A.~Ali, S.~Dutta, E.~N.~Saridakis and A.~A.~Sen,
  arXiv:1004.2474 [astro-ph.CO];
  S.~Dutta and E.~N.~Saridakis,
  JCAP {\bf 1005}, 013 (2010)
  [arXiv:1002.3373 [hep-th]];
  S.~Dutta and E.~N.~Saridakis,
  JCAP {\bf 1001}, 013 (2010)
  [arXiv:0911.1435 [hep-th]].
  

\bibitem{FT}
  G.~R.~Bengochea,
  arXiv:1008.3188 [astro-ph.CO];
 J.~B.~Dent, S.~Dutta and E.~N.~Saridakis,
  arXiv:1008.1250 [astro-ph.CO];
  P.~Wu and H.~Yu,
  Phys.\ Lett.\  B {\bf 692}, 176 (2010)
  [arXiv:1007.2348 [astro-ph.CO]];
  K.~K.~Yerzhanov, S.~R.~Myrzakul, I.~I.~Kulnazarov and R.~Myrzakulov,
  arXiv:1006.3879 [gr-qc];
  R.~Myrzakulov,
  arXiv:1006.1120 [gr-qc];
  P.~Wu and H.~Yu,
  arXiv:1006.0674 [gr-qc];
  E.~V.~Linder,
  Phys.\ Rev.\  D {\bf 81}, 127301 (2010)
  [arXiv:1005.3039 [astro-ph.CO]];
  G.~R.~Bengochea and R.~Ferraro,
  Phys.\ Rev.\  D {\bf 79}, 124019 (2009)
  [arXiv:0812.1205 [astro-ph]].



\bibitem{Perivolaropoulos:2005}
  L.~Perivolaropoulos,
  JCAP {\bf 0510}, 001 (2005)
  [arXiv:astro-ph/0504582].
 \bibitem{CaldwellCoorayMelchiorri}
  R.~Caldwell, A.~Cooray and A.~Melchiorri,
  Phys.\ Rev.\  D {\bf 76}, 023507 (2007)
  [arXiv:astro-ph/0703375].
 \bibitem{JainZhang}
  B.~Jain and P.~Zhang,
  arXiv:0709.2375 [astro-ph].
  \bibitem{NesserisPerivolaropoulos:2006}
  S.~Nesseris and L.~Perivolaropoulos,
  Phys.\ Rev.\  D {\bf 73}, 103511 (2006)
  [arXiv:astro-ph/0602053].
  \bibitem{WangHuiMayHaiman}
   S.~Wang, L.~Hui, M.~May and Z.~Haiman,
  Phys.\ Rev.\  D {\bf 76}, 063503 (2007)
  [arXiv:0705.0165 [astro-ph]].
  \bibitem{Tsujikawa}
   S.~Tsujikawa,
  Phys.\ Rev.\  D {\bf 76}, 023514 (2007)
  [arXiv:0705.1032 [astro-ph]].

   \bibitem{HeavensKitchingVerde}
   A.~F.~Heavens, T.~D.~Kitching and L.~Verde,
  arXiv:astro-ph/0703191;



\bibitem{Copeland}
E.J. Copeland, M. Sami, and S. Tsujikawa, Int. J. Mod. Phys. D
{\bf 15}, 1753 (2006).

\bibitem{HutererTurner}
  J.~Frieman, M.~Turner and D.~Huterer,
  Ann.\ Rev.\ Astron.\ Astrophys.\  {\bf 46}, 385 (2008)
  [arXiv:0803.0982 [astro-ph]].


\bibitem{LinderCahn:2007}
  E.~V.~Linder and R.~N.~Cahn,
  Astropart.\ Phys. {\bf{28}} (2007) 481

\bibitem{Gong:2009sp}
  Y.~Gong, M.~Ishak and A.~Wang,
  Phys.\ Rev.\  D {\bf 80}, 023002 (2009)
  [arXiv:0903.0001 [astro-ph.CO]].

\bibitem{Nesseris:2007pa}
  S.~Nesseris and L.~Perivolaropoulos,
  Phys.\ Rev.\  D {\bf 77}, 023504 (2008)
  [arXiv:0710.1092 [astro-ph]].

  \bibitem{dgp}
 C. Deffayet, Phys. Lett. B {\bf502}, 199 (2001).

\bibitem{Uzan:2006mf}
  J.~P.~Uzan,
  Gen.\ Rel.\ Grav.\  {\bf 39}, 307 (2007)
  [arXiv:astro-ph/0605313].

\bibitem{Polarski:2007rr}
  D.~Polarski and R.~Gannouji,
  arXiv:0710.1510 [astro-ph].

  \bibitem{Nesseris:2006er}
  S.~Nesseris and L.~Perivolaropoulos,
  JCAP {\bf 0701}, 018 (2007)
  [arXiv:astro-ph/0610092].

\bibitem{Hawkins:2002sg}
  E.~Hawkins {\it et al.},
  Mon.\ Not.\ Roy.\ Astron.\ Soc.\  {\bf 346}, 78 (2003)
  [arXiv:astro-ph/0212375];E.~V.~Linder,
  arXiv:0709.1113 [astro-ph].


  \bibitem{Viel:2004bf}
  M.~Viel, M.~G.~Haehnelt and V.~Springel,
  Mon.\ Not.\ Roy.\ Astron.\ Soc.\  {\bf 354}, 684 (2004)
  [arXiv:astro-ph/0404600].

\bibitem{Viel:2005ha}
  M.~Viel and M.~G.~Haehnelt,
  Mon.\ Not.\ Roy.\ Astron.\ Soc.\  {\bf 365}, 231 (2006)
  [arXiv:astro-ph/0508177].

  \bibitem{Kaiser:1996tp}
  N.~Kaiser,
  Astrophys.\ J.\  {\bf 498}, 26 (1998)
  [arXiv:astro-ph/9610120]; L.~Amendola, M.~Kunz and D.~Sapone,
  arXiv:0704.2421 [astro-ph];H.~Hoekstra {\it et al.},
  Astrophys.\ J.\  {\bf 647}, 116 (2006)
  [arXiv:astro-ph/0511089].

\bibitem{Seo:2003pu}
  H.~J.~Seo and D.~J.~Eisenstein,
  Astrophys.\ J.\  {\bf 598}, 720 (2003)
  [arXiv:astro-ph/0307460]; D.~Sapone and L.~Amendola,
  arXiv:0709.2792 [astro-ph].

\bibitem{Mantz:2007qh}
  A.~Mantz, S.~W.~Allen, H.~Ebeling and D.~Rapetti,
  arXiv:0709.4294 [astro-ph].

\bibitem{Pogosian:2005ez}
M. J.Rees and D. W. Sciama, Nature {\bf 217} 511 (1968); R. G.
Crittenden and N. Turok, Phys. Rev. Lett. {\bf 76}, 575 (1996);
  L.~Pogosian, P.~S.~Corasaniti, C.~Stephan-Otto, R.~Crittenden and R.~Nichol,
  Phys.\ Rev.\ D {\bf 72}, 103519 (2005)
  [arXiv:astro-ph/0506396].


  \bibitem{DentDutta:2008}
 J.~B.~Dent and S.~Dutta,
 Phys.\ Rev.\ D {\bf{79}} (2009) 063516
 [arXiv:0808.2689]

\bibitem{DentDuttaPerivolaropoulos:2009}
 J.~B.~Dent, S.~Dutta, and L.~Perivolaropoulos,
 Phys.\ Rev.\ D {\bf{80}} (2009) 023514
 [arXiv:0903.5296]

 \bibitem{Wenbinlin:2001}
  W.-B.~Lin,
  Chin\ .Phys.\ Lett.\  {\bf{18}} 1539 (2001)
  [astro-ph/0107162]


  \bibitem{faraoni}
 V. Faraoni, \textit{Cosmology in Scalar-Tensor Gravity}, vol. 139 of \textit{Fundamental Theories of Physics} (Kluwer Academic Publishers,
Dordrecht, 2004).

\bibitem{Will:1994fb}
  C.~M.~Will,
  Phys.\ Rev.\  D {\bf 50}, 6058 (1994)
  [arXiv:gr-qc/9406022].

\bibitem{Damour:1995kt}
  T.~Damour and G.~Esposito-Farese,
  Phys.\ Rev.\  D {\bf 53}, 5541 (1996)
  [arXiv:gr-qc/9506063].

\bibitem{Chiba:1999wt}
  T.~Chiba,
  Phys.\ Rev.\  D {\bf 60}, 083508 (1999)
  [arXiv:gr-qc/9903094].



\bibitem{EspositoFarese:2004cc}
  G.~Esposito-Farese,
  AIP Conf.\ Proc.\  {\bf 736}, 35 (2004)
  [arXiv:gr-qc/0409081].

\bibitem{Clifton:2004st}
  T.~Clifton, D.~F.~Mota and J.~D.~Barrow,
  Mon.\ Not.\ Roy.\ Astron.\ Soc.\  {\bf 358}, 601 (2005)
  [arXiv:gr-qc/0406001].

\bibitem{Ni:2005ej}
  W.~T.~Ni,
  Int.\ J.\ Mod.\ Phys.\  D {\bf 14}, 901 (2005)
  [arXiv:gr-qc/0504116].

\bibitem{Turyshev:2008dr}
  S.~G.~Turyshev,
  Ann.\ Rev.\ Nucl.\ Part.\ Sci.\  {\bf 58}, 207 (2008)
  [arXiv:0806.1731 [gr-qc]].

\bibitem{Tsujikawa:2008uc}
  S.~Tsujikawa, K.~Uddin, S.~Mizuno, R.~Tavakol and J.~Yokoyama,
  Phys.\ Rev.\  D {\bf 77}, 103009 (2008)
  [arXiv:0803.1106 [astro-ph]].

\bibitem{Chen:1999qh}
  X.~l.~Chen and M.~Kamionkowski,
  Phys.\ Rev.\  D {\bf 60}, 104036 (1999)
  [arXiv:astro-ph/9905368].

\bibitem{Damour:1993id}
  T.~Damour and K.~Nordtvedt,
  Phys.\ Rev.\  D {\bf 48}, 3436 (1993).


\bibitem{Santiago:1998ae}
  D.~I.~Santiago, D.~Kalligas and R.~V.~Wagoner,
  Phys.\ Rev.\  D {\bf 58}, 124005 (1998)
  [arXiv:gr-qc/9805044].


\bibitem{Damour:1998ae}
  T.~Damour and B.~Pichon,
  Phys.\ Rev.\  D {\bf 59}, 123502 (1999)
  [arXiv:astro-ph/9807176].


\bibitem{Navarro:1999ss}
  A.~Navarro, A.~Serna and J.~M.~Alimi,
  Phys.\ Rev.\  D {\bf 59}, 124015 (1999)
  [arXiv:astro-ph/9903368].





\bibitem{Perrotta:1999am}
  F.~Perrotta, C.~Baccigalupi and S.~Matarrese,
  Phys.\ Rev.\  D {\bf 61}, 023507 (2000)
  [arXiv:astro-ph/9906066].


\bibitem{Holden:1999hm}
  D.~J.~Holden and D.~Wands,
  Phys.\ Rev.\  D {\bf 61}, 043506 (2000)
  [arXiv:gr-qc/9908026].

\bibitem{Bartolo:1999sq}
  N.~Bartolo and M.~Pietroni,
  Phys.\ Rev.\  D {\bf 61}, 023518 (2000)
  [arXiv:hep-ph/9908521].

\bibitem{Gaztanaga:2000vw}
  E.~Gaztanaga and J.~A.~Lobo,
  Astrophys.\ J.\  {\bf 548}, 47 (2001)
  [arXiv:astro-ph/0003129].

\bibitem{Capozziello:2005mj}
  S.~Capozziello, S.~Nojiri and S.~D.~Odintsov,
  Phys.\ Lett.\  B {\bf 634}, 93 (2006)
  [arXiv:hep-th/0512118].

\bibitem{Gannouji:2006jm}
  R.~Gannouji, D.~Polarski, A.~Ranquet and A.~A.~Starobinsky,
  JCAP {\bf 0609}, 016 (2006)
  [arXiv:astro-ph/0606287].

\bibitem{Capozziello:2007iu}
  S.~Capozziello, S.~Nesseris and L.~Perivolaropoulos,
  JCAP {\bf 0712}, 009 (2007)
  [arXiv:0705.3586 [astro-ph]].

\bibitem{Demianski:2007mz}
  M.~Demianski, E.~Piedipalumbo, C.~Rubano and P.~Scudellaro,
  Astron.\ Astrophys.\  {\bf 481}, 279 (2008)
  [arXiv:0711.1043 [astro-ph]].

\bibitem{Barenboim:2007bu}
  G.~Barenboim and J.~Lykken,
  JCAP {\bf 0803}, 017 (2008)
  [arXiv:0711.3653 [astro-ph]].

\bibitem{Jarv:2008eb}
  L.~Jarv, P.~Kuusk and M.~Saal,
  Phys.\ Rev.\  D {\bf 78}, 083530 (2008)
  [arXiv:0807.2159 [gr-qc]].


\bibitem{Vitagliano:2009zy}
  V.~Vitagliano, S.~Liberati and V.~Faraoni,
  Class.\ Quant.\ Grav.\  {\bf 26}, 215005 (2009)
  [arXiv:0906.5429 [gr-qc]].

 \bibitem{Tatsuya:2009}
  T. ~Narikawa, and K.~Yamamoto,
  [arXiv:0912.1445[astro-ph.CO]]

\bibitem{Song:2010}
 Y.-S.~Song, L.~Hollenstein, G.~Caldera-Cabral, and K.~Koyama,
 [arXiv:1001.0969 [astro-ph.CO]]














\bibitem{DiPorto:2007ym}
  C.~Di Porto and L.~Amendola,
  arXiv:0707.2686 [astro-ph].





\bibitem{Bertschinger:2006aw}
  E.~Bertschinger,
  Astrophys.\ J.\  {\bf 648}, 797 (2006)
  [arXiv:astro-ph/0604485].

\bibitem{Gong:2008fh}
  Y.~Gong,
  Phys.\ Rev.\  D {\bf 78}, 123010 (2008)
  [arXiv:0808.1316 [astro-ph]].






\bibitem{Dodelson:2003ft}
  S.~Dodelson,
{\it  Amsterdam, Netherlands: Academic Pr. (2003) 440 p}

\bibitem{Damour:1996ke}
  T.~Damour and G.~Esposito-Farese,
  Phys.\ Rev.\  D {\bf 54}, 1474 (1996)
  [arXiv:gr-qc/9602056].
  
\bibitem{bagram}
  S.~Baghram and S.~Rahvar,
  arXiv:1004.3360 [astro-ph.CO].
  
\bibitem{weller}
  S.~A.~Appleby and J.~Weller,
  arXiv:1008.2693 [astro-ph.CO].




\bibitem{Sanchez:2010ng}
  J.~C.~B.~Sanchez and L.~Perivolaropoulos,
  arXiv:1002.2042 [Unknown].





\end{thebibliography}
\end{document}